\documentclass[manuscript]{aastex}
\usepackage{graphics,color}
\usepackage{epsfig}
\usepackage{amssymb}
\usepackage{times}

\newcommand{\lum}{erg\, s$^{-1}$}

\newcommand{\Zsun}{Z$_\odot$}
\newcommand{\Msun}{\ensuremath{M_{\odot}}}

\begin{document}
\title{Investigation of X-ray cavities in the cooling flow system Abell 1991}
\shorttitle{X-ray cavities in Abell 1991}
\shortauthors{M. B.Pandge et al.}
\author{M.B.Pandge, N.D.Vagshette, S.S.Sonkamble, M.K.Patil\altaffilmark{*}}
\affil{School of Physical Sciences, Swami Ramanand Teerth Marathwada University, Nanded-431606, Maharashtra India \\ e-mail: patil@iucaa.ernet.in}
\email{patil@iucaa.ernet.in}
\begin{abstract}
We present results based on the systematic analysis of \textit{Chandra} archive data on the X-ray bright Abell Richness class-I type cluster Abell 1991 with an objective to investigate properties of the X-ray cavities hosted by this system. The unsharp masked image as well as 2-d $\beta$ model subtracted residual image of Abell 1991 reveals a pair of X-ray cavities and a region of excess emission in the central $\sim$12 kpc region. Both the cavities are of ellipsoidal shape and exhibit an order of magnitude deficiency in the X-ray surface brightness compared to that in the undisturbed regions. Spectral analysis of X-ray photons extracted from the cavities lead to the temperature values equal to $1.77_{-0.12}^{+0.19}$ keV for N-cavity and $1.53_{-0.06}^{+0.05}$ keV for S-cavity, while that for the excess X-ray emission region is found to be equal to $2.06_{-0.07}^{+0.12}$ keV. Radial temperature profile derived for Abell 1991 reveals a positive temperature gradient, reaching to a maximum of 2.63 keV at $\sim$ 76~kpc and then declines in outward direction. 0.5$-$2.0~keV soft band image of the central 15\arcsec region of Abell 1991 reveals relatively cooler three different knot like features that are about 10\arcsec off the X-ray peak of the cluster. Total power of the cavities is found to be equal to $\sim 8.64\times 10^{43}$ \lum, while the X-ray luminosity within the cooling radius is found to be 6.04 $\times 10^{43}$ \lum, comparison of which imply that the mechanical energy released by the central AGN outburst is sufficient to balance the radative loss. 
\end{abstract}
\section{Introduction}
\begin{figure*}
\centering
\includegraphics[width=65mm,height=65mm]{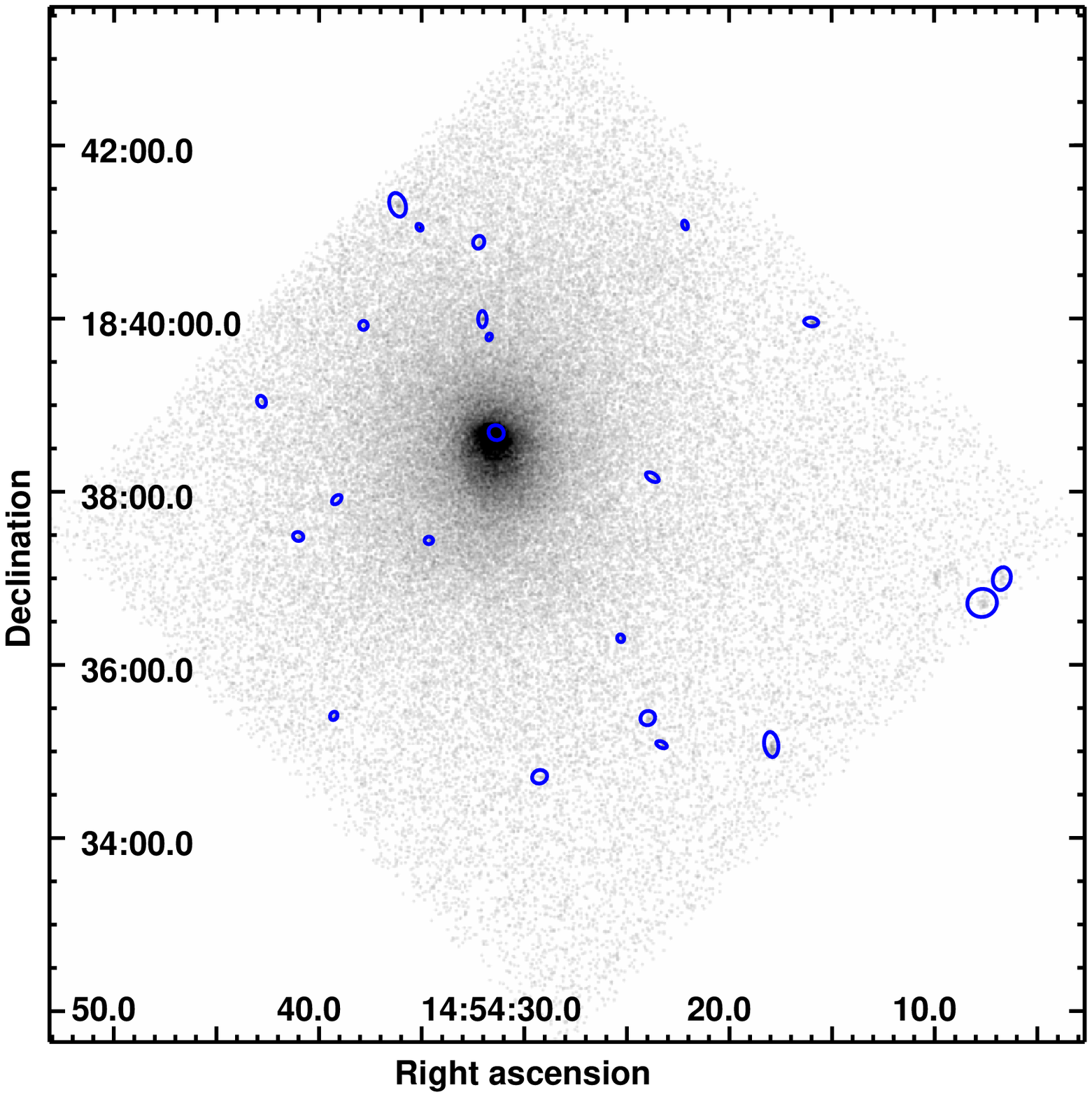}
\caption{\label{overlay.fig} 0.5-7.0 keV raw {\it Chandra} ACIS-S3 image of Abell 1991, overlaid on which are the 21 point sources detected within the chip S3.}
\end{figure*} 
High-resolution X-ray observations of bright cool-core galaxy clusters with \textit{Chandra} and \textit{XMM-Newton} telescopes have shown that, even in the cores of clusters with high cooling rate, the observed quantum of gas cooled below T $\le$ 1\,keV is significantly small \citep{2011ApJ...728...54Z}. Further, the rate at which mass is being deposited in the cores of the dominant cooling flow clusters is found to be an order of magnitude smaller than that estimated using standard models \citep{2003ApJ...590..207P, 2006PhR...427....1P}. This means some compensatory steady heating mechanism must be operative in the cores of such clusters \citep{2006ApJ...648..164M}. A variety of mechanisms have been proposed to compensate the radiative losses, that include, the magnetic field re-connection (\cite{1990ApJ...348...73S}), thermal conduction due to electron collisions \citep{2001ApJ...562L.129N}, turbulent conduction \citep{2004MNRAS.347.1130V} and heating by cosmic rays \citep{2008A&A...484...51C}. However, the most promising source of heating for replenishing the radiative loss of the hot gas in clusters and groups of galaxies is the energy input from the super-massive black hole (SMBH) that is residing at the core of such cluster \citep{1993MNRAS.264L..25B, 1995MNRAS.276..663B, 2006ApJ...648..164M, 2006MNRAS.366..417F}. One of the early clues explaining role of the active galactic nucleus (AGN) in reheating of the inter-galactic medium (IGM) was the high detection rate ($\sim$70\%) of radio activity in the central galaxy of the X-ray bright clusters relative to the X-ray faint clusters \citep{1990AJ.....99...14B}.

X-ray deficient cavities detected in numerous groups and clusters of galaxies provided with the direct evidence for the AGN feedback in such systems \citep{2004ApJ...607..800B,2009ApJ...705..624D,2010ApJ...712..883D,2011ApJ...726...86R}. Often such cavities are seen near the center of clusters in about few to few tens of kpc around the X-ray center and appear as buoyantly rising bubbles in the intra-cluster medium (ICM) \citep{2010ApJ...712..883D, 2009ApJ...705..624D, 2012MNRAS.421..808P}. Further, such cavities are found in association with radio jets or bubbles / lobes and are often filled with relativistic particles and magnetic fields \citep{2007ARA&A..45..117M,2012AdAst2012E...6G}. Wealth of the observational evidences over last few decades have shown that, roughly 70 - 75\% of the X-ray bright cool-core clusters harbor detectable number of cavities with energy input from the central AGN roughly sufficient to compensate the cooling of gas in the clusters \citep{2006ApJ...652..216R}. This means, feedback from the central AGN plays an important role in shaping the morphology of the hot gas and also in the thermodynamical evolution of the galaxy clusters \citep{2008MNRAS.384..251G}, providing with a reliable means to estimate the mechanical energy injected by the SMBH in to the ICM \citep{2005ApJ...624..586J}. 

In this paper we present results based on the analysis of \textit{Chandra} observations of the X-ray bright cluster Abell 1991 with an objective to investigate  properties of the X-ray cavities hosted by this system. Abell 1991 is classified as the Bautz-Morgan type I \citep{1970BAAS....2R.294B} cluster with a centrally dominant galaxy and is defined as the Abell Richness Class I type system \citep{1999ApJS..125...35S}. Structure of the paper is as follows: Section 2 describes  X-ray observations and data preparation methodology. Section 3 discusses the imaging analysis with an emphasis on the investigation of X-ray cavities. In Section 4 we discuss spectral properties of hot gas within X-ray cavities and other regions of interest, while Section 5 discusses energetics associated with the central engine and correspondence between the X-ray and radio observations. Summary of the results from the study is presented in Section 6. We assume $H_0$=73 km s$^{-1}$ Mpc$^{-1}$ in this paper. At the red-shift of Abell 1991, luminosity distance is 255 Mpc and corresponding angular size is 1.17 kpc arcsec$^{-1}$. 

\section[Observations]{Observations and Data Preparation}
\label{obs} 
\begin{figure*}
\vbox
{
\includegraphics[width=82mm,height=70mm]{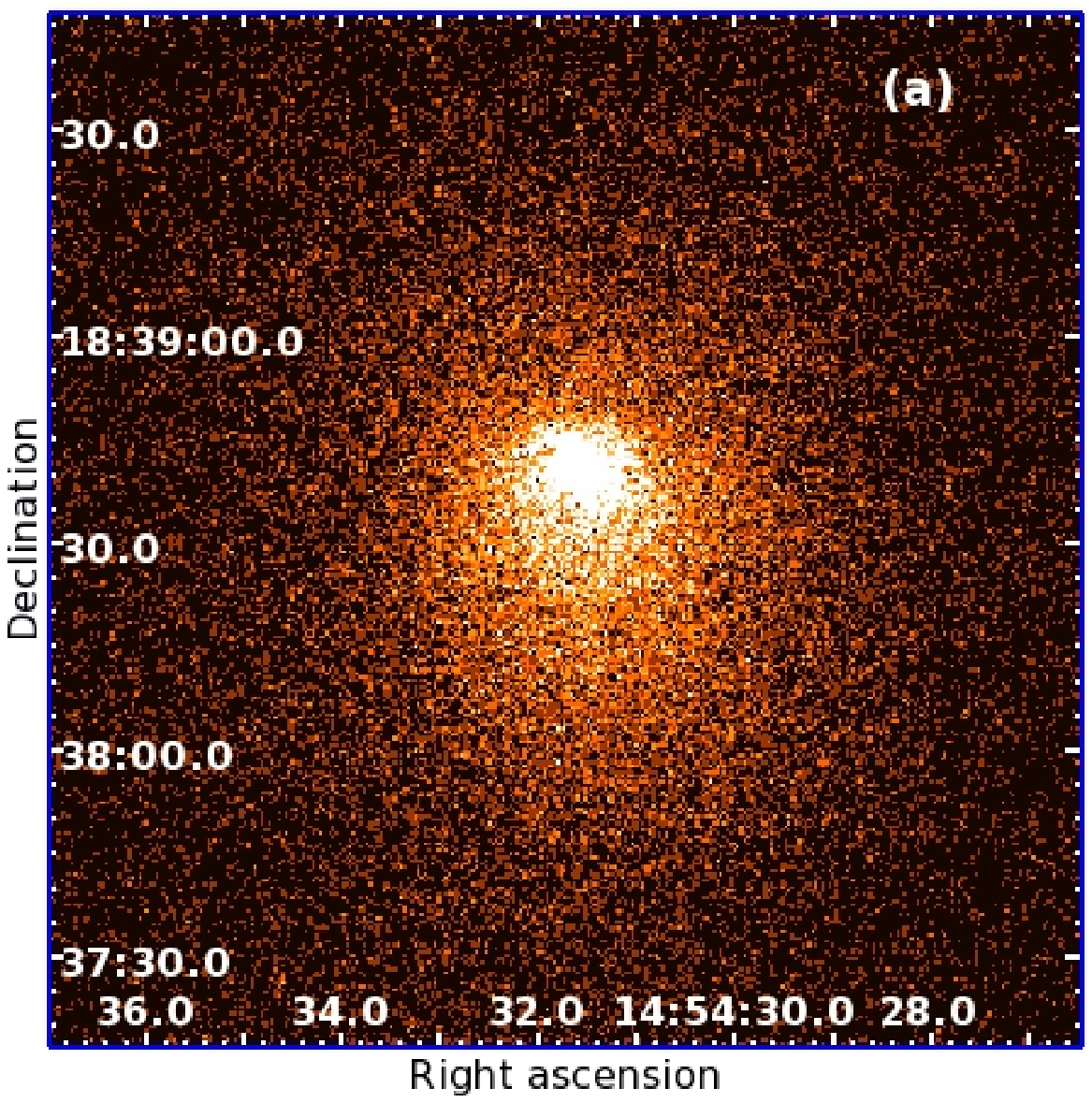}\hskip+3mm
\includegraphics[width=82mm,height=70mm]{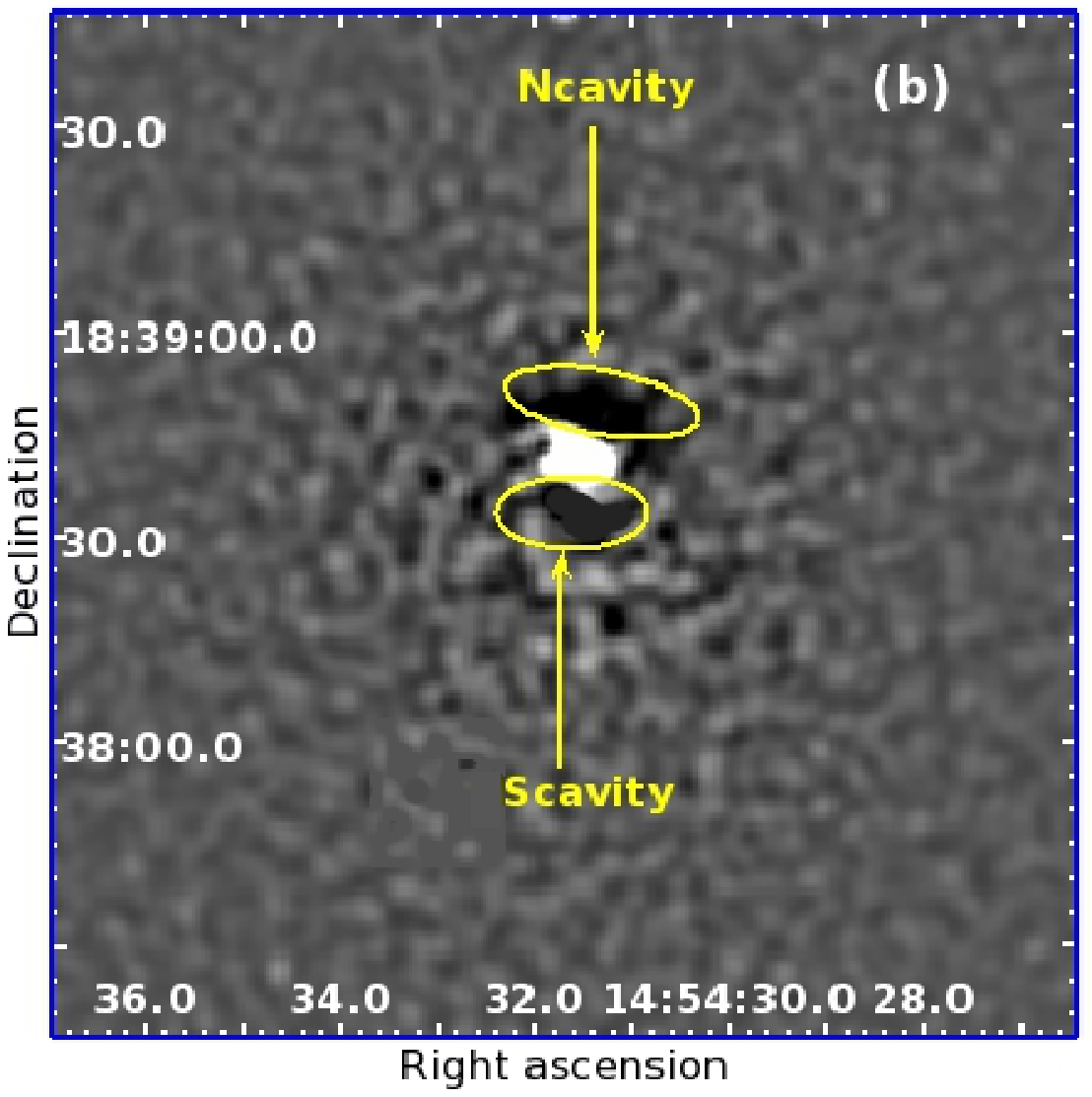}\hskip-3mm
\includegraphics[width=82mm,height=70mm]{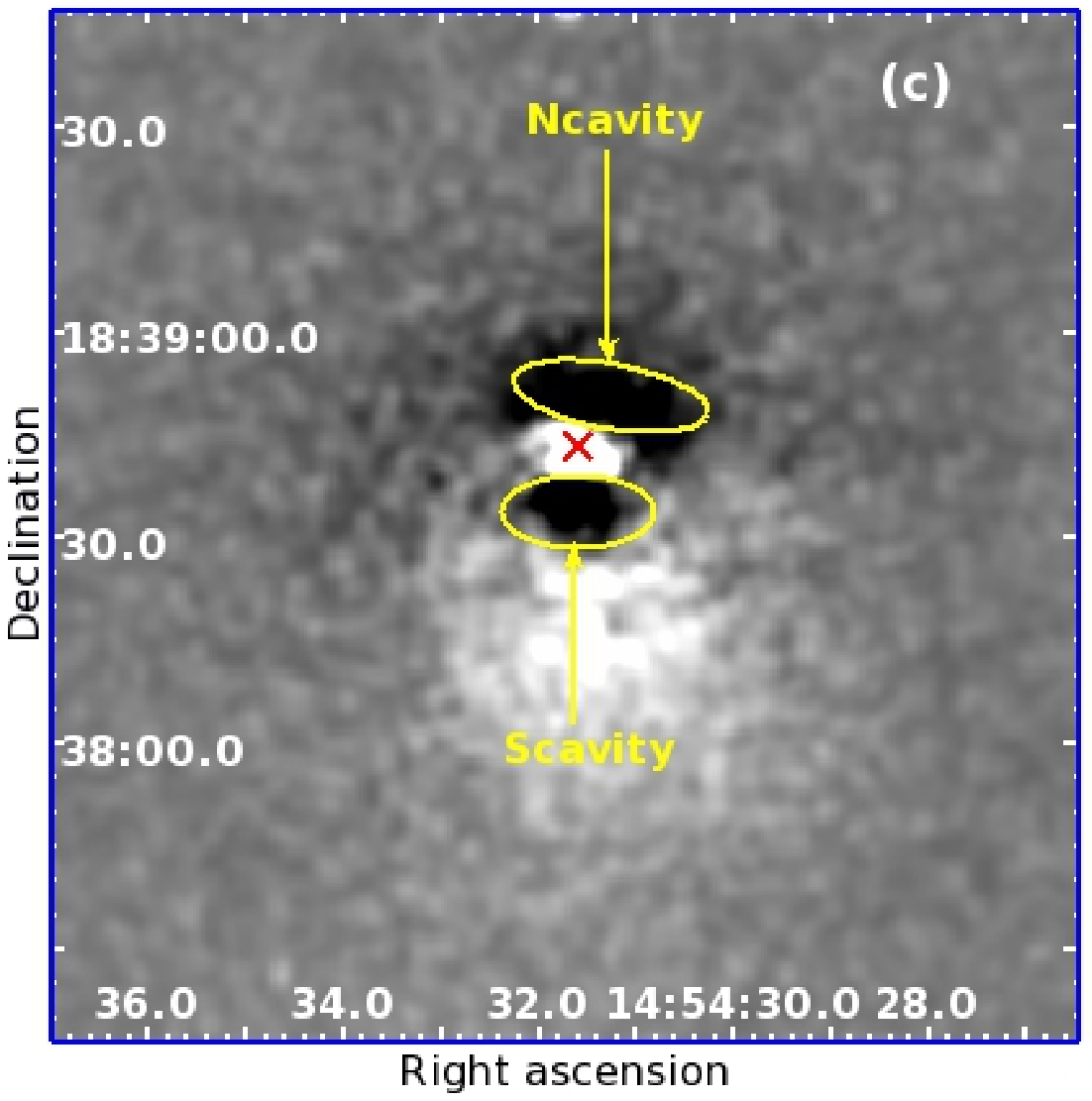}\hskip+3mm
\includegraphics[width=82mm,height=70mm]{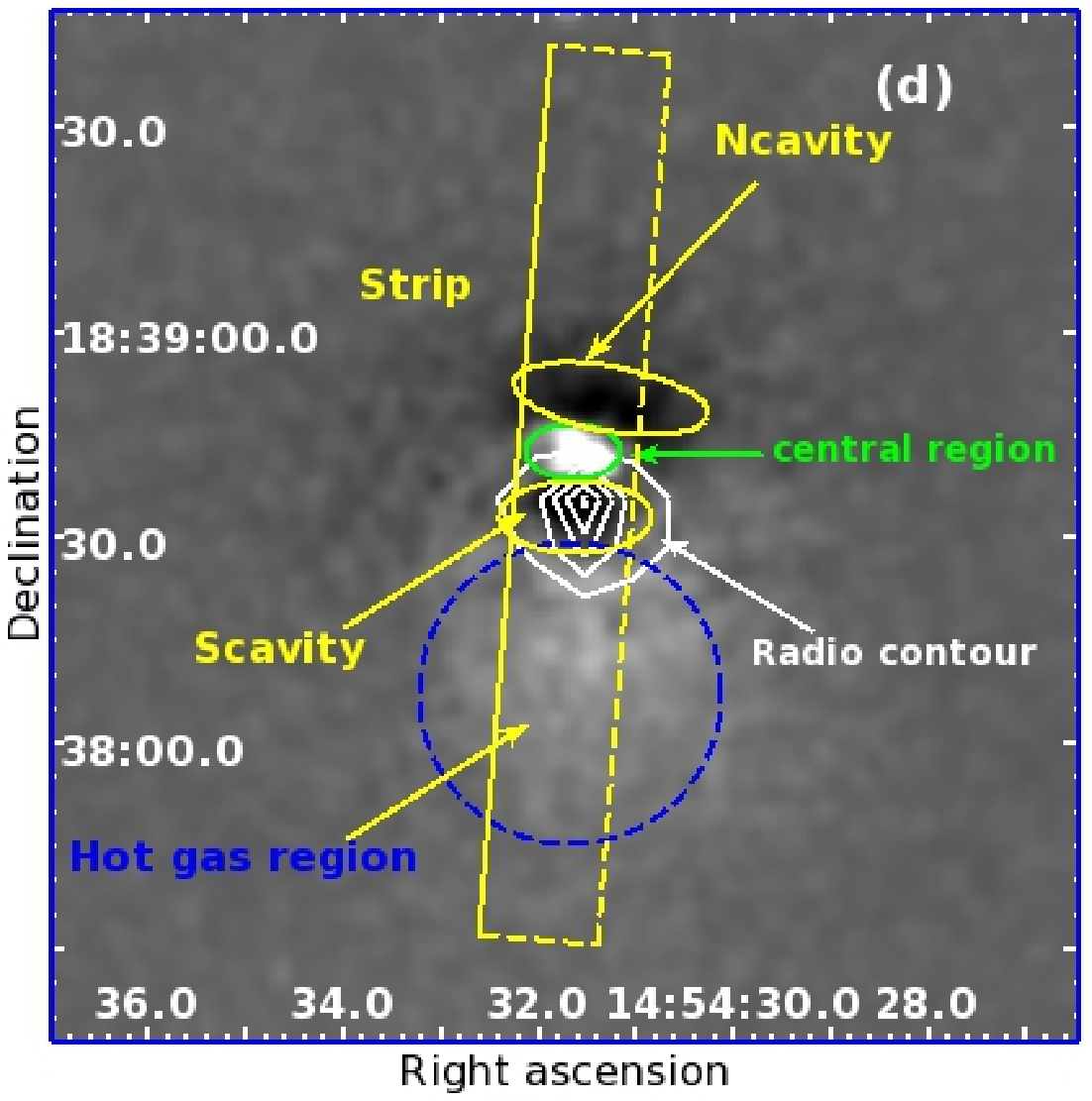}\hskip+3mm
}
\caption{(a) ACIS-S3 exposure corrected, background subtracted 0.5-3.0\,keV \emph{Chandra} image of Abell 1991. (b) \emph{Chandra} 0.5-3.0\,keV unsharp masked image of Abell 1991 derived after subtracting a 5$\sigma$ wider Gaussian kernel smoothed image from that smoothed with a narrow 3$\sigma$ Gaussian kernel. This figure reveals substructures in the central region. (c) Elliptical 2-d beta model subtracted residual image exhibiting a pair of X-ray cavities and an excess X-ray emission region. (d) Same as that in (c) overlaid with VLA FIRST survey radio contours (white). The radio contours are at 0.5, 7.3, 14, 21, 27.23 mJy/beam levels, image rms is $\sim$ 145 $\mu$Jy. Regions of interest are also shown in this figure.}
\label{ecimage}
\end{figure*}

Abell 1991 was observed with  \textit{Chandra Advanced CCD Imaging Spectrometer} (ACIS-S) on 16-17 December 2002 (ObsID 3193, see \cite{2004ApJ...613..180S} for details) for a total exposure of 38.8 ks. Event-2 data sets available in the archive of \textit{Chandra} observatory were acquired for the present study and were reduced following the standard tasks available within the X-ray analysis package \textit{CIAO} V 3.4.0\footnote{Chandra Interactive Analysis Observation package} in conjunction with the  \textit{Chandra} calibration data base (\textit{CALDB}) V 3.4.0 provided by the \textit{Chandra} X-ray Center\footnote{http://cxc.harvard.edu/}. Initially the event files were screened for cosmic rays using ASCA grades and were reprocessed to apply most up-to-date corrections for the time-dependent gain change, charge transfer efficiency, and degraded quantum efficiency of the ACIS detector. Though particle background during observations was stable and no flares have been detected \citep{2004ApJ...613..180S}, however, as a routine process we performed light curve filtering to detect the time intervals affected by  the background counts exceeding 20\% of their mean count rate. This resulted in the net effective exposure time of 38.3\,ks. X-ray background component was adequately modeled using the ``blank-sky" data sets\footnote{http://cxc.harvard.edu/contrib/maxim/acisbg}, that included both particle-induced and unresolved sky components. Spectra from regions of interest during this study were extracted using the {\ttfamily ACISSPEC} task and corresponding Redistribution Matrix Files (RMF), Ancillary Response Files (ARF) were generated. The point source detection was performed on CCD ID 7 using {\tt CIAO} {\ttfamily WAVDETECT} task with the source detection threshold set to 10$^{-6}$. This enabled us to detect a total of 21 point sources within S3 chip . We generated exposure corrected, background subtracted 0.5 - 3.0 keV energy band image using {\ttfamily DMMERGE} task available within {\tt CIAO} 3.4.0 and is shown in Fig.~\ref{ecimage}(a). This figure reveals the bright X-ray region with its peak emission centroid located at (RA = 14:54:31.7) and (DEC = +18:38:42), showing an offset of $\sim$ 10\arcsec relative to that in optical/radio image (RA = 14:54:31.5; DEC= +18:38:32) centered on NGC 5778 \citep{1999ApJS..125...35S}.

\section{X-ray Imaging Analysis} 
\subsection{X-ray Cavity Detection}
\label{Cavity Detection} 
A pair of X-ray cavities in Abell 1991 has already been reported by \citet{2010ApJ...712..883D}. However, with an objective to enhance visualization and to investigate properties of the X-ray cavities in this system, we employed a variety of image processing techniques. These includes, deriving unsharp masked image and residual image of Abell 1991. Figure~\ref{ecimage} (b) shows 0.5-3.0\,keV  \emph{Chandra}  soft band unsharp masked image of Abell 1991 derived after subtracting a 5$\sigma$ wider Gaussian kernel smoothed image from that smoothed with a narrow 3$\sigma$ Gaussian kernel (see \cite{2010ApJ...712..883D,2012MNRAS.421..808P} for details). Resulting unsharp masked image reveals a pair of X-ray cavities, one on the Northern side (N-cavity) and other on the Southern side (S-cavity) of the X-ray center of Abell 1991. Both these cavities are roughly of ellipsoidal shape and are located at a projected distance of $\sim$ 12.0 kpc from the X-ray center of the Abell 1991. The BCG appears to coincide with the S-cavity and shows an off-set of 10\arcsec relative to the X-ray peak. As S-cavity has associated radio source, therefore, is defined as  ``clear cavity", while N-cavity failing to show such an association, can be defined as ``ghost cavity" \citep{2001ApJ...562L.149M}. In addition to the pair of prominent cavities, an excess X-ray emission is also evident on the Southern side of the S-cavity. 

Features evident in unsharp masked image were further confirmed using residual map produced after subtracting 2-d smooth model of the  Abell 1991 from its original image. A 2-d smooth model of Abell 1991 was generated by fitting $\beta$ model to the clean, background subtracted image using fitting function \textit{Sherpa} available within \textit{CIAO}. Fitting parameters i.e., ellipticity, position angle, normalization angle, and local background, etc. were kept free during this process. The best fit 2-d model was then subtracted from the cleaned image of Abell 1991 to produce its residual map and is shown in Figure~\ref{ecimage} (c). This figure confirms a pair of ellipsoidal cavities evident in the unsharp masked image and exhibit surface brightness fluctuations more clearly. 

\begin{figure*}
\hbox
{
\includegraphics[width=80mm,height=80mm]{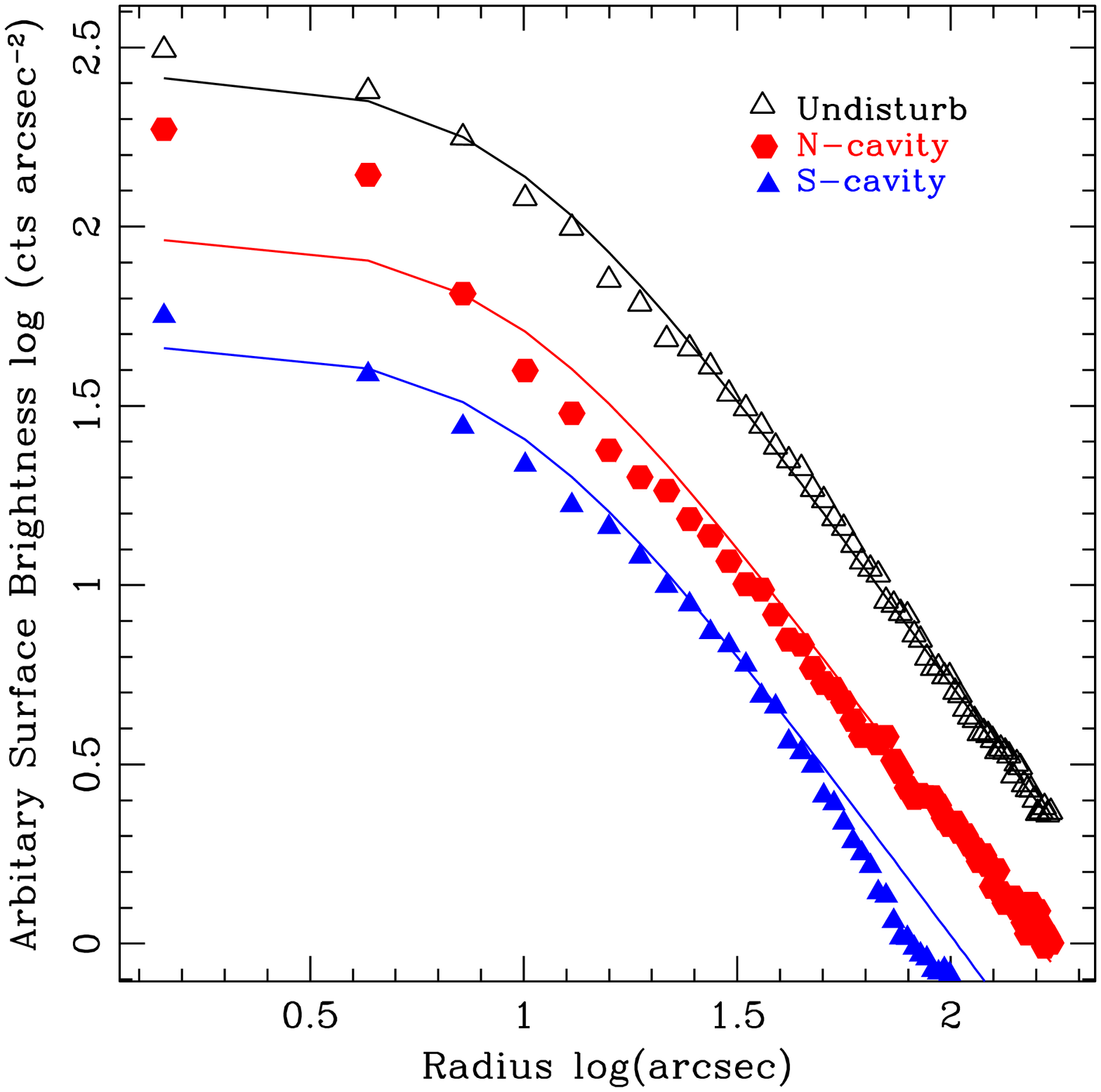}
\includegraphics[width=80mm,height=80mm]{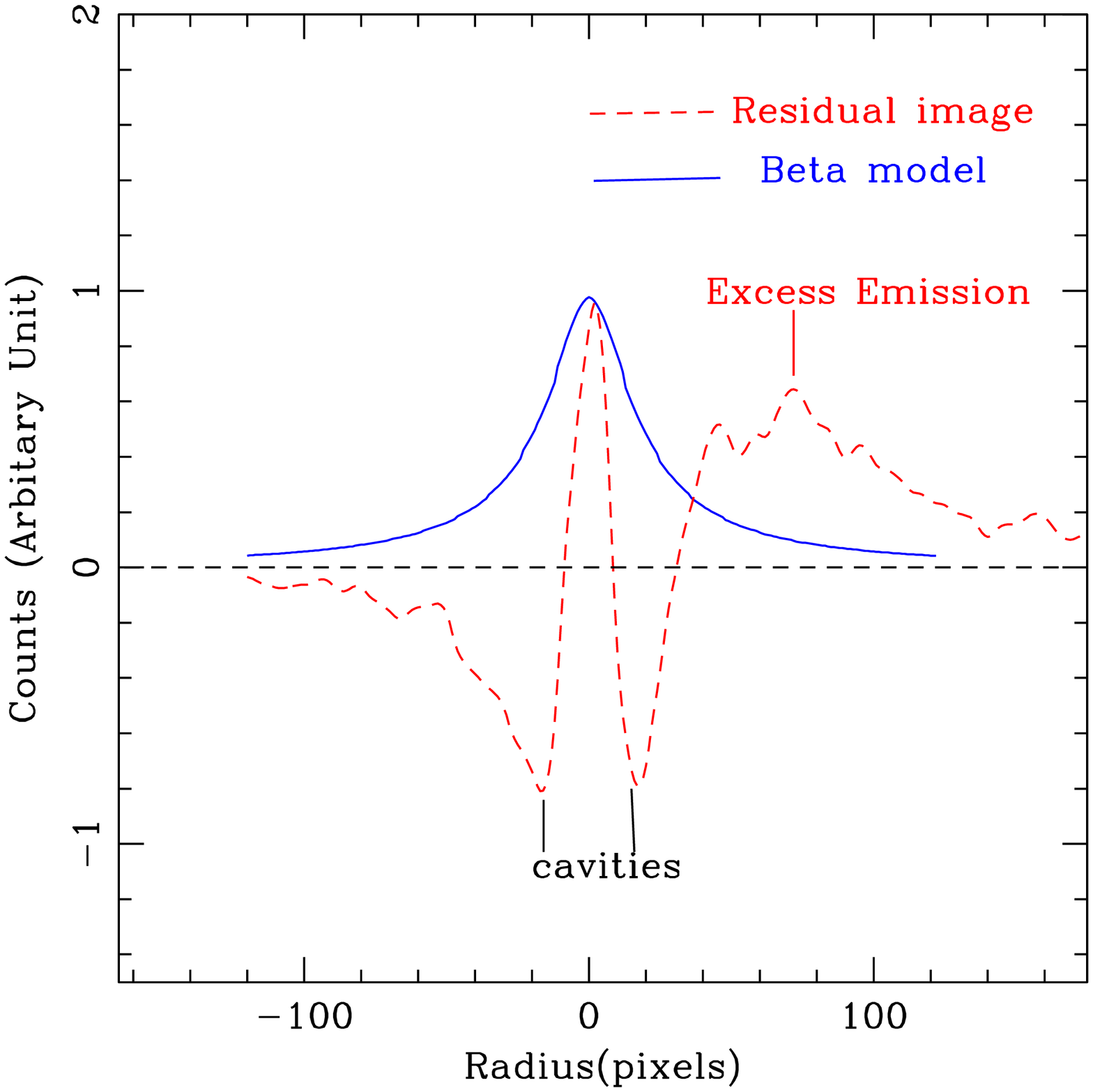}
}
\caption{\label{sur} {\it (left panel):} 0.5-5.0 keV azimuthally averaged surface brightness profile of the undisturbed X-ray photon distribution (open triangles) within Abell 1991. Continuous line in it represents the best fit $\beta$-model. We neglected the regions of X-ray cavities as well as excess emission during this fit. Surface brightness profiles of the X-ray photons extracted along the conical sectors 0-140$^o$ covering N-cavity (red hexagon) and 220-230$^o$ covering S-cavity (blue filled triangles) are also shown in this figure. The profile extracted along 220-230$^o$ conical sector also reveals an excess emission at $\sim$ 40\arcsec. {\it (right panel):} X-ray photon distribution within the 0.5-5.0 keV background subtracted 2d $\beta$-model image (blue solid line) and those in the 2-d $\beta$ model subtracted residual image (red dashed line). Significant depressions at the locations of both the cavities and an excess emission at $\sim$ 20\arcsec -70\arcsec are clearly evident in this figure.}
\end{figure*}

\subsection{Azimuthally Averaged Surface Brightness Profile}
Unsharp masked image as well as the 2-d $\beta$ model subtracted residual image revealed a pair of X-ray cavities and an excess X-ray emission region (Fig.~\ref{ecimage} b $\&$ c). To examine the extent of flux variation in these regions, we derived surface brightness profile of Abell 1991 using  \textit{Chandra} ACIS-S3 data. For this an azimuthally averaged surface brightness profile was computed over 0.5-5.0\,keV range by extracting X-ray photons form concentric circular annuli each of width equal to 2.5\arcsec starting from 2\arcsec up to 175\arcsec. To produce profile of the ``undisturbed" ICM, we excluded the cavity regions during this process. The resultant surface brightness profile was fitted with a single $\beta$-model using the tool \texttt{SHERPA} with the $\chi^2$ statistics of Gehrels variance \citep{1986ApJ...303..336G}, 
 \[\Sigma(r)=\Sigma_0\left[ 1+\left( \frac{r}{r_0} \right)^2\right]^{-3\beta+0.5}\]

where $\Sigma(r)$ is the X-ray brightness at the projected distance r, $\Sigma_0$ is the central surface brightness, $r_{0}$ is the core radius of the X-ray emission and $\beta$ represent the slope of the surface brightness profile. Like in other cooling flow systems, distribution of the ICM in Abell 1991 shows an excess emission in the central region relative to the best fitted model (Fig.~\ref{sur} {\it left panel}), whose parameters are r$_c=14.50\pm0.15$ kpc and $\beta=0.44\pm0.01$.

To examine the extent of fluctuations in the surface brightness distribution due to the presence of cavities, we considered two conical sectors along the cavity directions i.e.,  between 0-140$^o$ and 220-360$^o$ for N and S cavities, respectively, and extracted X-ray photons from 40 different annuli along these sectors. Plots of the count rates along these sectors as a function of radial distance are presented in Figure~\ref{sur} ({\it left panel}).The profiles derived for both the sectors reveal a significant drop in the surface brightness distribution between 8\arcsec and 15\arcsec, roughly of the cavity sizes, compared with the best-fit model. Additionally, an excess emission is also evident in the profile extracted along the S-cavity and corresponds to the bright plume like feature shown by blue dotted circle in Figure~\ref{ecimage}(d). To enhance the fluctuations seen in the regions of interest, we also plot the X-ray photon distribution extracted from the rectangular strip region of the residual map  (Figure~\ref{ecimage}-d) and is shown by red dashed curve in Figure~\ref{sur} {\it right panel}). For comparison we also show the photon distribution extracted from similar region in the 2-d smooth model image (blue solid line). This figure clearly reveals a significant fraction of fluctuations in the X-ray photon count rates across the cavity positions in the residual image compared to the smooth distribution in the model image. An excess X-ray emission is also evident at about 40\arcsec and corresponds to the bright plum region in Figure~\ref{ecimage}(d).

\begin{figure}
\includegraphics[width=70mm,height=70mm]{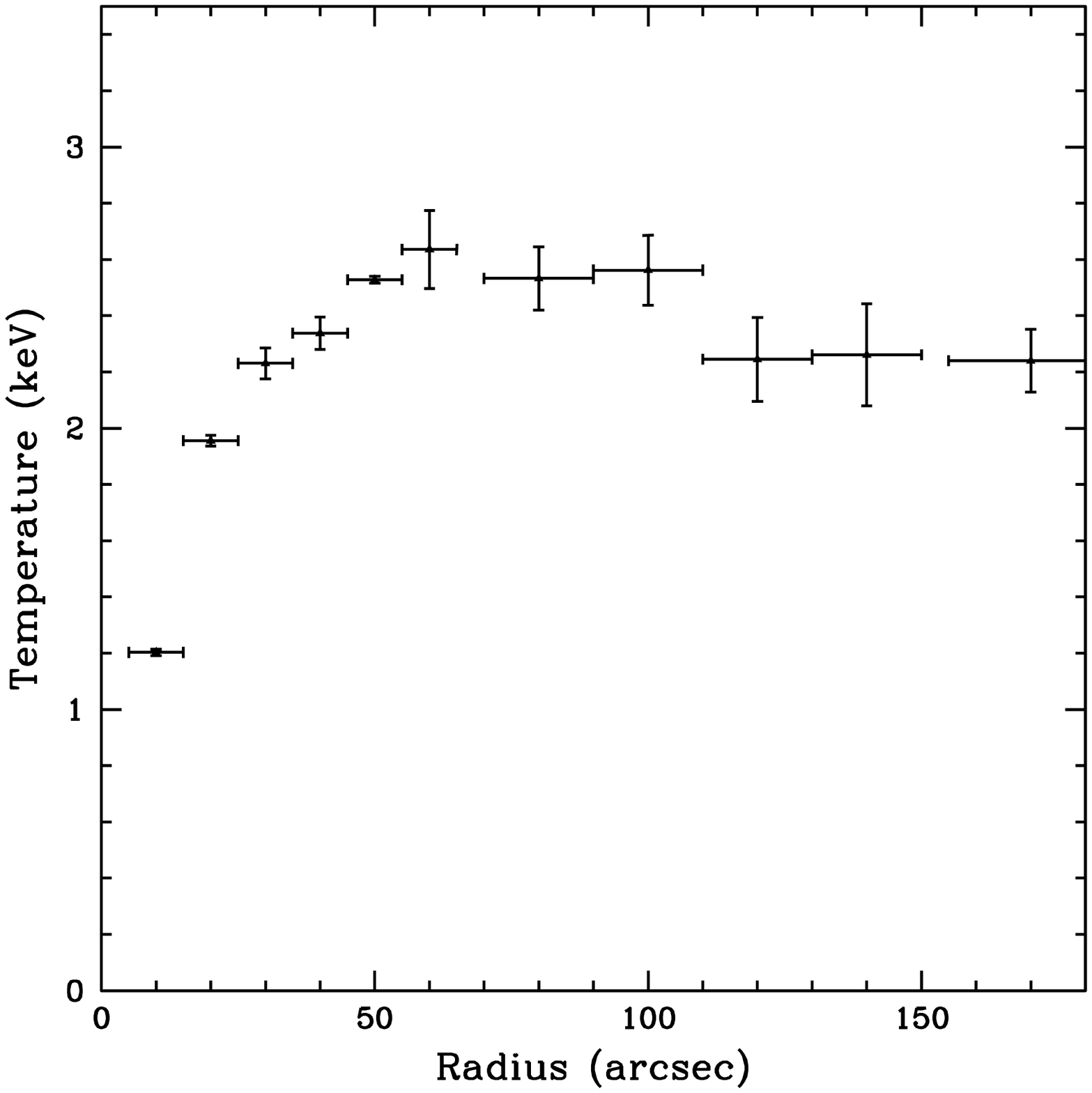}
\centering
\caption{0.5-7.0 keV radial temperature profile of the hot gas distribution in Abell 1991. This was obtained by fitting a single-temperature \textit{apec} model to the spectra extracted form 11 different annuli centered on the X-ray peak of Abell 1991 cluster. Errorbars show the 90\% confidence level.}
\label{projtemp}
\end{figure}
\section{\textit{Chandra} Spectral Analysis} 
A significant amount of variation, particularly in the central region,  was evident in the surface brightness distribution of the Abell 1991  (Fig.~\ref{ecimage}). With an objective to examine their relevance with temperature variation, we performed spectral analysis of the X-ray photons extracted from different regions of interest. For this, 0.5-7.0 keV X-ray spectra were extracted from each of the region using the \textit{CIAO} task \texttt{specextract} and were grouped such as to get at least 20 counts in each bin. During this study we fixed the hydrogen column density at the Galactic value (N$_H$ = $2.46 \times 10^{20} cm^{-2}$) \citep{1990ARA&A..28..215D} and the redshift at $z$ = 0.0587. The metal abundance and temperature were allowed to vary during the fit. 
\begin{figure}
\centering
\includegraphics[width=80mm,height=80mm]{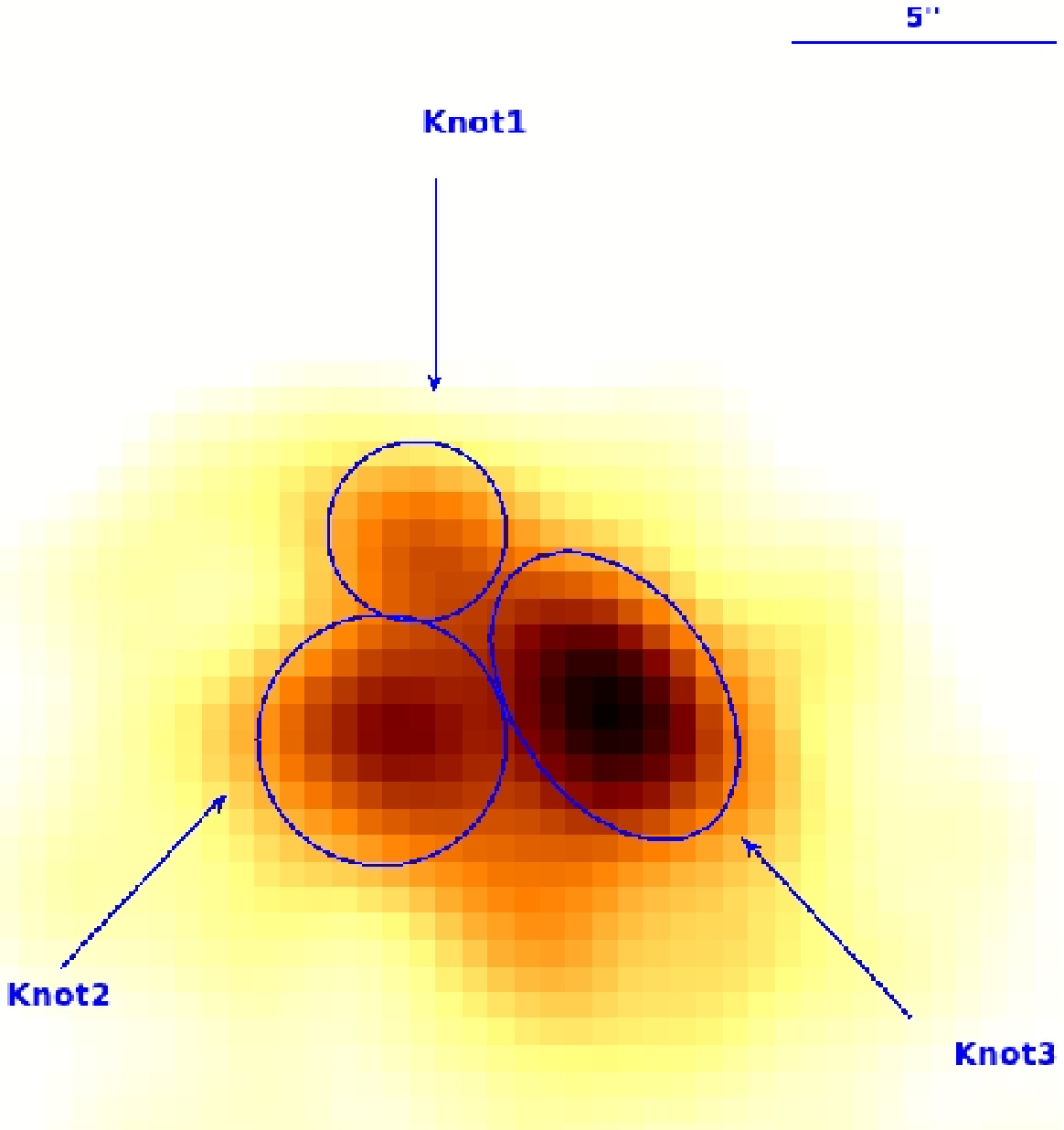}
\caption{ \label{knots} 0.5-2.0 keV exposure corrected background subtracted adaptively smoothed {\it Chandra} image of central 15\arcsec region of Abell 1991. This figure clearly revels three different knots in central region. Spectra representative from each of the knot have been extracted separately.} 
\end{figure}
\begin{table*} 
\caption{Spectral properties of X-ray cavities, central knots, diffuse component (hot gas) and the central source}
\begin{tabular}{@{}cccccccr@{}} 
\hline 
{\it Regions}  			 &$N_H^{intr}$         		   & Best-fit 		& $kT$   	        & $kT$        & Abund	&$\chi^{2}$/dof \\ 
         			 &($10^{20}\,cm^{-2}$) 		   & Model    		&(KeV)  	        & (KeV)       &  \Zsun     	& \\ 
\hline 
\textit{N-cavity}                & $-$       			  &(phabs*apec)            &$1.77_{-0.12}^{+0.19}$ &$-$ &$0.47_{-0.15}^{+0.25}$ &47/62 \\
\textit{S-cavity}                & $-$        			  &(phabs*apec)            &$1.53_{-0.06}^{+0.05}$ &$-$ &$0.48_{-0.10}^{+0.11}$ &178/137 \\ 
\textit{Excess emission reg.}    & $-$         		          &(phabs*apec)            &$2.06_{-0.07}^{+0.12}$ &$-$ &$0.61_{-0.12}^{+0.09}$ &180/156 \\
\textit{Central source}          &$6.95_{-0.03}^{+0.02}$          &(phabs*phabs*apec+pow)  &$0.93_{-0.02}^{+0.03}$ &$-$ &$0.37_{-0.05}^{+0.07}$ &70.20/75 \\
\textit{Total diff emission}     &$-$			          &(phabs*apec+apec)       &$1.96_{-0.22}^{+0.04}$ &$6.84_{-3.94}^{+2.06}$ &$0.83_{-0.3}^{+0.3}$ &372/241 \\ 
\textit{Three knots} 		 &$7.50_{-0.02}^{+0.02}$          &(phabs*phabs*apec)      &$0.94_{-0.03}^{+0.03}$ &$-$ &$0.25_{-0.04}^{+0.05}$ &54/66 \\
\hline 
\end{tabular}
\footnotesize \begin{flushleft} {\bf Note} { Col. 1 - regions of interest used to extract the spectra in the energy band 0.5-7.0\,keV, Col. 2 - hydrogen column density intrinsic to each of the source region, Col. 3 - best-fit model, Col. 4 \& 5 - low and high temperature values of the best fitted spectra, Col. 6 - metallicity values and Col.7 - Goodness of fit.} 
\end{flushleft} 
\label{specpar} 
\end{table*}

To determine the global spectral properties of diffuse gas in Abell 1991, we extracted a combined spectrum from within 180\arcsec ($\sim$ 210\,kpc), excluding  central 5\arcsec region and the resolved point sources. A total of 85,844 background-subtracted counts were extracted in the range between 0.5$-$7.0\,keV. Resultant spectrum was then fitted with a single temperature (\texttt{phabs*apec}) model which resulted in $\chi^2=582$ for $312$ {\it dof}. Careful look at the residuals revealed that the single temperature model is not adequate to constrain the observed spectrum. Therefore, we tried with a double temperature model \texttt {phabs(apec+apec)}, which improved the fit to $\chi^2/dof =372/241$. The best fit lower and higher temperature values are found to be $\sim$ $1.96_{-0.22}^{+0.04}$~keV and $\sim$ $6.84_{-3.94}^{+2.06}$~keV, respectively, for the two components and the corresponding X-ray luminosity is found to be $7.65\times 10^{43} erg \, {s}^{-1}$ (Table ~\ref{specpar}). \citet{2004ApJ...613..180S} fitting double temperature (\texttt{MEKAL}) model to the X-ray photons extracted from within 190\arcsec arrive at temperature values equal to 1.51$\pm$0.09 keV and $7.87\pm 0.96$ keV. Our estimate of lower temperature value is higher than that derived by \cite{2004ApJ...613..180S}, while the higher temperature value shows a reasonable agreement. \citet{1984ApJ...285....1S} form deprojection analysis of total spectrum over Einstein 0.5-3.0 keV could find the temperature kT$\sim$1.6 keV. 

\subsection{Azimuthally Averaged Spectral Analysis} 
A projected radial temperature profile was also derived for Abell 1991 by analyzing spectra extracted from 11 different annuli centered on the cluster core. Widths of the annuli were adjusted such as to achieve roughly same S/N on each of the annulus. The background-subtracted 0.5-7.0\,keV spectrum extracted from each of the annuls was then fitted with an \texttt{apec} thermal plasma component modified by the intrinsic absorption (\texttt{phabs*apec}). We kept temperature and metallicity free during this fit. Profile of the best fitted temperature along with associated errors is shown in Fig.~\ref{projtemp}. Like in other cooling flow galaxies, temperature profile derived for Abell 1991 also shows a positive temperature gradient in the inner region, reaches to a maximum of 2.63 keV at about $\sim$76~kpc and then declines in the outward direction, consistent with that reported by \cite{2004ApJ...613..180S} and \cite{2009ApJS..182...12C}.
 
\subsection{Cavities and excess emission region} 
To perform a more detailed analysis of the temperature structure of the hot gas corresponding to the cavities and the excess X-ray emission regions, we extracted spectra representative of both the cavities (shown by yellow ellipses in Fig.~\ref{ecimage} d) as well as from the excess X-ray emission region (shown by dotted blue region in Fig.~\ref{ecimage}d). 0.5-7.0\,keV spectra from these regions were then fitted independently with the single temperature model (\texttt{phabs*apec}) within {\tt XSPEC}, with the column density set to the Galactic value and keeping temperature and abundance values free. This yielded the best fit temperature values for N and S cavities to be equal to $1.77_{-0.12}^{+0.19}$ keV and $1.53_{-0.06}^{+0.05}$ keV, respectively, while that for the excess emission region was found to be $2.06_{-0.07}^{+0.12}$ \,keV. The metal abundance associated with the excess emission is found to be 0.61$^{-0.12}_{+0.09}$ \Zsun. This means the region of excess emission is hotter compared to the cavity regions, which themselves are relatively hotter than the undisturbed gas surrounding the cavities. It is likely that the X-ray photons extracted from cavities may also include photons from the foreground and/or background of the cavities. However, our projected and deprojected analysis of the X-ray photons from this system, particularly in the cavity locations, does not show significant variations in the estimated values of temperature as well as luminosity. Therefore, we believe that the effect of such emission, if any, would be negligible. 
  
\subsection{Central Source} 
To examine spectral properties of X-ray emission from the nuclear region of Abell 1991, we extracted a 0.5-7.0\,keV spectrum from the central 5\arcsec region centered on the X-ray peak of the cluster. This spectrum was fitted with a single temperature \texttt{apec} model, however, it resulted in poor fit particularly at higher energy range. Therefore, we added a power-law component to this model. The best-fit temperature obtained from the double component model is $0.93_{-0.02}^{+0.03}$ keV and is much cooler than the surrounding ICM, while the abundance is found to be equal to 0.37$^{-0.05}_{+0.07}$ \Zsun. The power-law index in the present case is $\Gamma$=1.4, indicative of the sufficient hard structure of the central source. The unabsorbed 0.5-7.0\,keV flux of the central source was found to be 5.13$\times$10$^{-13}$ erg s$^{-1}$ cm$^{-2}$ leading to a luminosity of 3.95$\times$ 10$^{42} erg \, {s}^{-1}$. The best fitted spectral parameters of the central source are given in Table~\ref{specpar}.

\subsection{X-ray knots} 
Exposure corrected, background subtracted adaptively smoothed 0.5-2.0\,keV \textit{Chandra} image revealed three different knot-like features in the central 15\arcsec region of the Abell 1991 that are about 10\arcsec off the X-ray peak (Fig.~\ref{knots}) and are consistent with those reported by \cite{2004ApJ...613..180S}. To examine properties of these knots, we  extracted a combined  spectrum from these  knots and exported it to \texttt{XSPEC} for spectral fitting. We fitted the spectrum with  an \texttt{apec} model including an additional absorption component. The best fit resulted in the  temperature equal to 0.94$\pm$0.03 keV and metallicity equal of $0.25_{-0.04}^{+0.05}$ \Zsun (Table~\ref{specpar}), while those by \cite{ 2004ApJ...613..180S} with  (\texttt{WABS+MEKAL}) are 0.83$\pm$0.02 keV and $0.46_{-0.08}^{+0.22}$ \Zsun, respectively. Our estimates are slightly higher than those of \cite{ 2004ApJ...613..180S}. Spectra representative from each of the three knots were also extracted and fitted separately. Spectra of knots 1 and 2 were constrained with simple (\texttt{apec}) component, however, that for knot 3 required an addition absorption component.

\begin{figure}
\centering
\includegraphics[width=80mm,height=80mm]{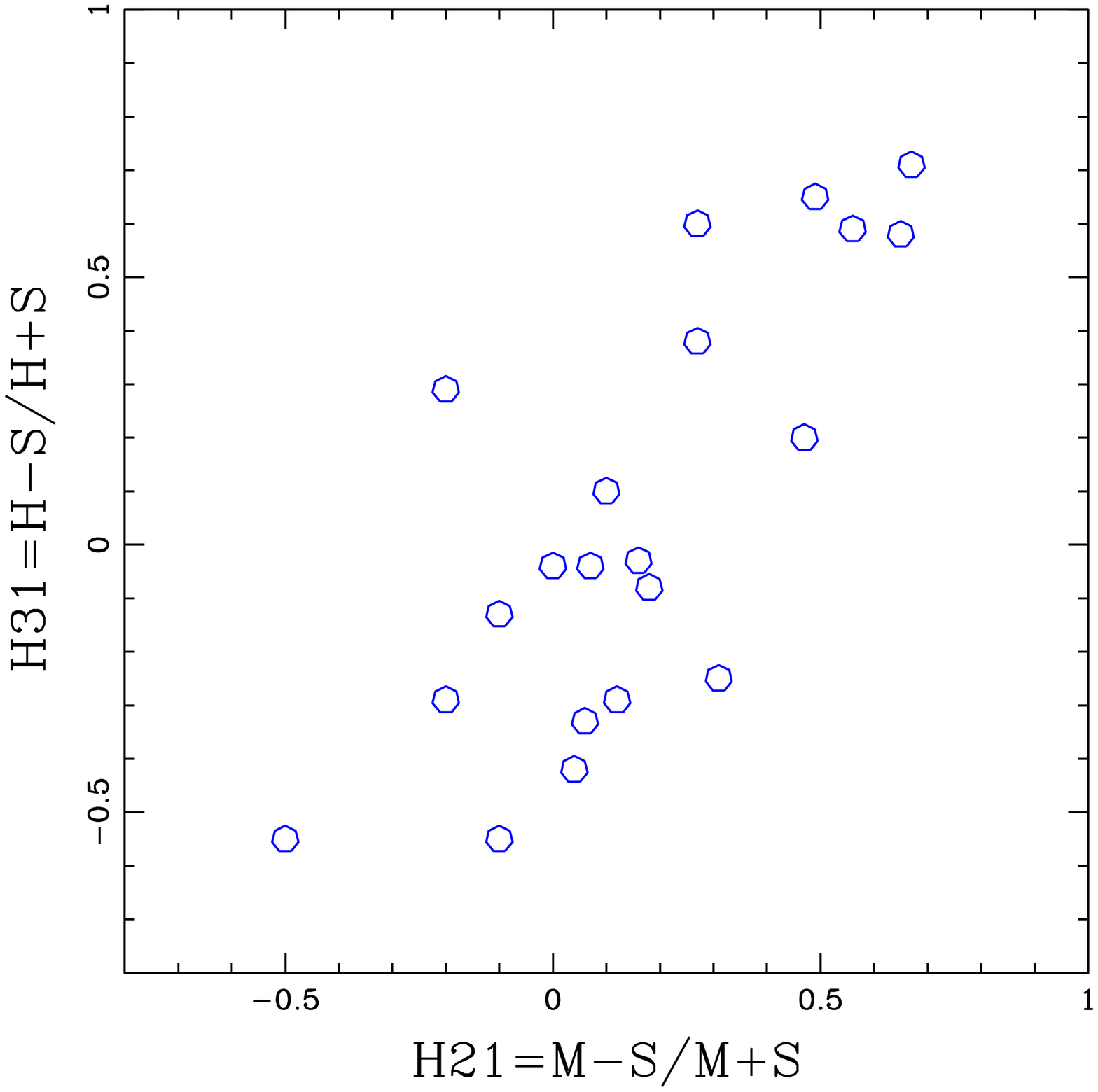}
\caption{\label{xcolor}  X-ray color-color plot of all the 21 point sources detected within 0.5$-$7.0 keV ACIS-S3 chip.} 
\end{figure}
\begin{table*} 
\caption{Properties of the point sources detected within ACIS-S3 chip}
\begin{tabular}{@{}cccccccr@{}} 
\hline 
No. & RA & DEC &Area &H21 &H31 & Count Rate &Luminosity~~~ \\ 
& & &$(arcsec^{2})$ & & & $(10^{-3} counts~{s}^{-1})$ &($10^{40} erg~s^{-1}$) \\
\hline 
1&14:54:41.017 & +18:37:29.06 &34.86 & 0.47 & 0.20    & 0.548  & 3.51~~~~~~~~~~  \\ 
2 &14:54:39.282 & +18:35:24.72 &24.21 & 0.65 & 0.58   & 0.992  & 6.35~~~~~~~~~~  \\ 
3 &14:54:39.137 & +18:37:54.61 &30.98 & 0.27 & 0.60   & 0.705  & 4.51~~~~~~~~~~   \\
4 &14:54:37.835 & +18:39:55.36 &28.08 & 0.67 & 0.71   & 0.627  & 4.01~~~~~~~~~~   \\ 
5 &14:54:35.101 & +18:41:03.35 &15.49 & -0.12 &-0.55  & 2.611  & 1.67~~~~~~~~~~   \\ 
6 &14:54:32.035 & +18:39:59.72 &14.52 & 0.56 &0.59    & 1.305  & 8.35~~~~~~~~~~  \\ 
7 &14:54:31.704 & +18:39:47.34 &13.56 & 0.04 &-0.42   & 1.436  & 9.19~~~~~~~~~~   \\ 
8 &14:54:25.296 & +18:36:18.49 &20.33 & 0.31 &-0.25   & 0.913  & 5.85~~~~~~~~~~   \\ 
9 &14:54:22.154 & +18:41:04.92 &19.37 & 0.16&-0.03    & 1.384  & 8.86~~~~~~~~~~   \\ 
10 &14:54:42.809 & +18:39:02.64 &38.73 & -0.29 &-0.29 & 0.600  & 3.84~~~~~~~~~~   \\ 
11 &14:54:36.171 & +18:41:18.91 &99.73 & -0.52 &-0.55 & 1.593  & 1.02~~~~~~~~~~   \\ 
12 &14:54:34.645 & +18:37:26.32 &25.17& 0.07 &-0.04   &1.1230  & 7.19~~~~~~~~~~   \\ 
13 &14:54:32.214 & +18:40:52.96 &56.16 & -0.10 &-0.13 & 1.149  & 7.35~~~~~~~~~~   \\ 
14 &14:54:23.975 & +18:35:23.19 &77.46 & 0.00 &-0.04  & 0.913  & 5.85~~~~~~~~~~   \\ 
15 &14:54:23.306 & +18:35:04.75 &25.17 & 0.12 &-0.29  & 0.809  & 5.18~~~~~~~~~~   \\ 
16 &14:54:17.959 & +18:35:02.00 &94.89 & 0.49 & 0.65  & 2.454  & 1.57~~~~~~~~~~   \\ 
17 &14:54:16.001 & +18:39:57.66 &44.54 & 0.06 &-0.33  & 1.514  & 9.69~~~~~~~~~~   \\ 
18 &14:54:29.249 & +18:34:42.54 &80.37 & 0.27 & 0.38  & 1.044  & 6.68~~~~~~~~~~   \\ 
19 &14:54:23.751 & +18:38:10.14 &39.70 & 0.10 & 0.10  & 0.809  & 5.18~~~~~~~~~~   \\ 
20 &14:54:07.675 & +18:36:42.88 &29.05 & 0.18 &-0.08  & 0.600  & 3.84~~~~~~~~~~   \\ 
21 &14:54:06.714 & +18:36:59.70 &84.24 & -0.25 & 0.29 & 0.887  & 5.68~~~~~~~~~~   \\ 
\hline 
\hline 
\end{tabular}
\footnotesize \begin{flushleft} {\bf Notes:} { Col. 2 \& 3 - position coordinates of the resolved sources, Col. 4 - area information for each of the source, the hardness ratios H21 and H31 are given in the Col. 5 \& 6, respectively. Col. 7 \& 8- count rates and X-ray luminosities ($L_{X}$) of individual source, respectively.} 
\end{flushleft} 
\label{sources} 
\end{table*}

\subsection{X-ray Point Sources} 
The wavelet based detection algorithm {\ttfamily wavdetect}  within (\textit{CIAO}) enabled us to detect a total of 21 discrete sources within ACIS-S3 chip of 0.5 -7.0 \,keV \textit{Chandra} image of Abell 1991. As flux of the X-ray photons from individual  source was not enough to perform spectroscopy of individual source, therefore, we derived X-ray colors of these sources with an objective to investigate their structural properties \citep{2009ApJ...705.1632P,2012arXiv1205.6057V}. For this we extracted X-ray counts for each of the source in three different energy bands: soft (S, 0.5-1.0\,keV); medium (M, 1.0-2.0\,keV) and hard  (H, 2.0-7.0\,keV) using the task {\ttfamily dmextract}. Then the hardness ratios of the individual source were derived using definitions H21=(M-S)/(M+S) and H31=(H-S)/(H+S) and are listed in Table~\ref{sources}. We then plotted the X-ray color-color plot for the resolved sources and is shown in Figure~\ref{xcolor}, where are the sources are found to lie along the diagonal region covering hardness ratios in the range between [H21,H31] = [-0.55, -0.52] to [+0.68, +0.71]. Among them, four are harder sources with hardness ratios $>$ [+0.50, +0.50] and probably represent strongly absorbed AGNs \citep{2004ApJ...600..729R}.

\section{Discussion}
\subsection{Cavity Energetics}
\label{eng}
Comparison of the cavity power with the X-ray luminosity of the ICM within cooling radius allow us to examine balance between the radiative loss  from the ICM and the cavity heating \citep{2004ApJ...607..800B,2005MNRAS.364.1343D,2006ApJ...652..216R}. Therefore, an attempt was made to estimate power of both the cavities by measuring $pdV$ work done by the jets in inflating these cavities and by estimating their ages  \citep{2007ARA&A..45..117M}. To investigate properties of X-ray cavities and to quantify energetics of the AGN outburst it is required to know the ICM density, temperature and pressure at the cavity location. For this we used azimuthally averaged temperature profile of the X-ray photons derived above in $\S$4.1. Energetics of both the cavities were then determined following standard method and using the relation $E_{\rm cavity} = \frac{\gamma_1}{\gamma_1 - 1} pV$ \citep{2004ApJ...607..800B, 2006ApJ...652..216R}, where $p$ is thermal pressure of the ICM at the cavity positions and $V$ is the volume of each cavity. For the relativistic fluid using $\gamma_1$ = 4/3 this equation becomes $E_{\rm cavity} = 4pV$. Volumes of both the cavities were estimated assuming that they are symmetric about the plane of the sky with centers lying in the plane perpendicular to the line-of-sight and passing through the central AGN. Then the expression for volume of the cavity becomes $V = 4\pi a b R_{\rm los}$/3, where $a$ and $b$ are the projected semi-major and semi-minor axes of the cavities, respectively,  and R$_{\rm los}$ is the line-of-sight distance between the nuclear source and the center of each cavity. The measured dimensions of both the cavities along with their volumes are given in Table~\ref{cavity}. 
 
\begin{table} 
\centering
\caption{Properties of the X-ray cavities within Abell 1991} 
\begin{tabular}{@{}lccr@{}} 
\hline
\hline {\it Cavity Parameters} &{\it Ncavity} & {\it Scavity} \\ 
\hline $a$ (kpc) &16.8 & 13.3 \\ 
$b$ (kpc) &5.5& 6.2 \\ 
$R$ (kpc) &12.4 & 11.5 \\ 
$t_{c_s}$ (yr) &1.8$\times 10^7$ & 1.8$\times10^7$ \\ 
$t_{buoy}$(yr) &3.8$\times 10^7$ & 2.9$\times 10^7$ \\ 
$t_{refill}$(yr) &6.8$\times 10^7$ & 5.9$\times 10^7$ \\ 
$P_{cavity}$ (erg s$^{-1}$) &3.8$\times 10^{43}$ & 4.8$\times 10^{43}$ \\ 
\hline 
\end{tabular} 

\footnotesize
\begin{flushleft} {\bf Note} : a , b - semi major and semi minor axis lengths of the cavities, R projected distance from the center of the  cluster.
 \end{flushleft} 
\label{cavity}
\end{table}
 
 We adopted all the three methods proposed by \citet{2004ApJ...607..800B} to estimate ages of the cavities, namely,  (1) the cavity reaches the projected distance R from the center of the cluster in sound crossing time $t_s = R/c_s = R/\sqrt{\gamma kT/\mu m_p}$; where $\gamma$=4/3, $k$ is Boltzmann constant, T is temperature, $\mu$ is mean molecular weight and $m_p$ is mass of the proton; (2) cavities are rising buoyantly to the present position in time $t_{buoy} = R/\sqrt{2gV/SC}$, where $g=GM/R^2$ is the gravitational acceleration at the cavity position R, V is volume of the cavity, S is the cross-section of the cavity, and C = 0.75 is the drag coefficient; and (3) time required for the gas to refill the displaced volume of the cavity as it rises in the time given by $t_{ref} \sim 2\sqrt{r/g}$, where $r$ is radius of the cavity.  

The buoyancy rise time is the time taken by the cavity to reach its terminal velocity that depend on the drag forces and  in the case of clearly detached cavities provides better estimate of the age. The sound crossing time is the time taken by the bubble to travel the projected distance from central AGN to its present position at the sound speed, while the refill time is the time taken by a bubble to rise buoyantly through its own diameter starting from the rest. Typically, the ages estimated by all the three assumptions agree within a factor of 2 \citep{2004ApJ...607..800B}, with the buoyancy time lying between the sound crossing time and the refill time. We estimated ages of both the cavities using all the three methods discussed above and are given in Table~\ref{cavity}. Though all the three estimates do not vary significantly, which among these represent the best estimate of age is still not clear. Therefore, in the present case we took average of all the three estimates and derived the cavity power using relation $ P_{\rm cavity} = \frac{E_{\rm cavity}}{<t_{\rm age}>}$. Resultant values of total powers for the N and S-cavities in the present study are found to be equal to 3.8$\times 10^{43}$ erg\,s$^{-1}$ and 4.8$\times 10^{43}$ erg\,s$^{-1}$, respectively, and are given in Table~\ref{cavity}. As these estimates do not include possible hydrodynamical shocks, therefore, provides the lower limit to the total power of the AGN.

\begin{figure*}
\hbox
{
\includegraphics[width=80mm,height=80mm]{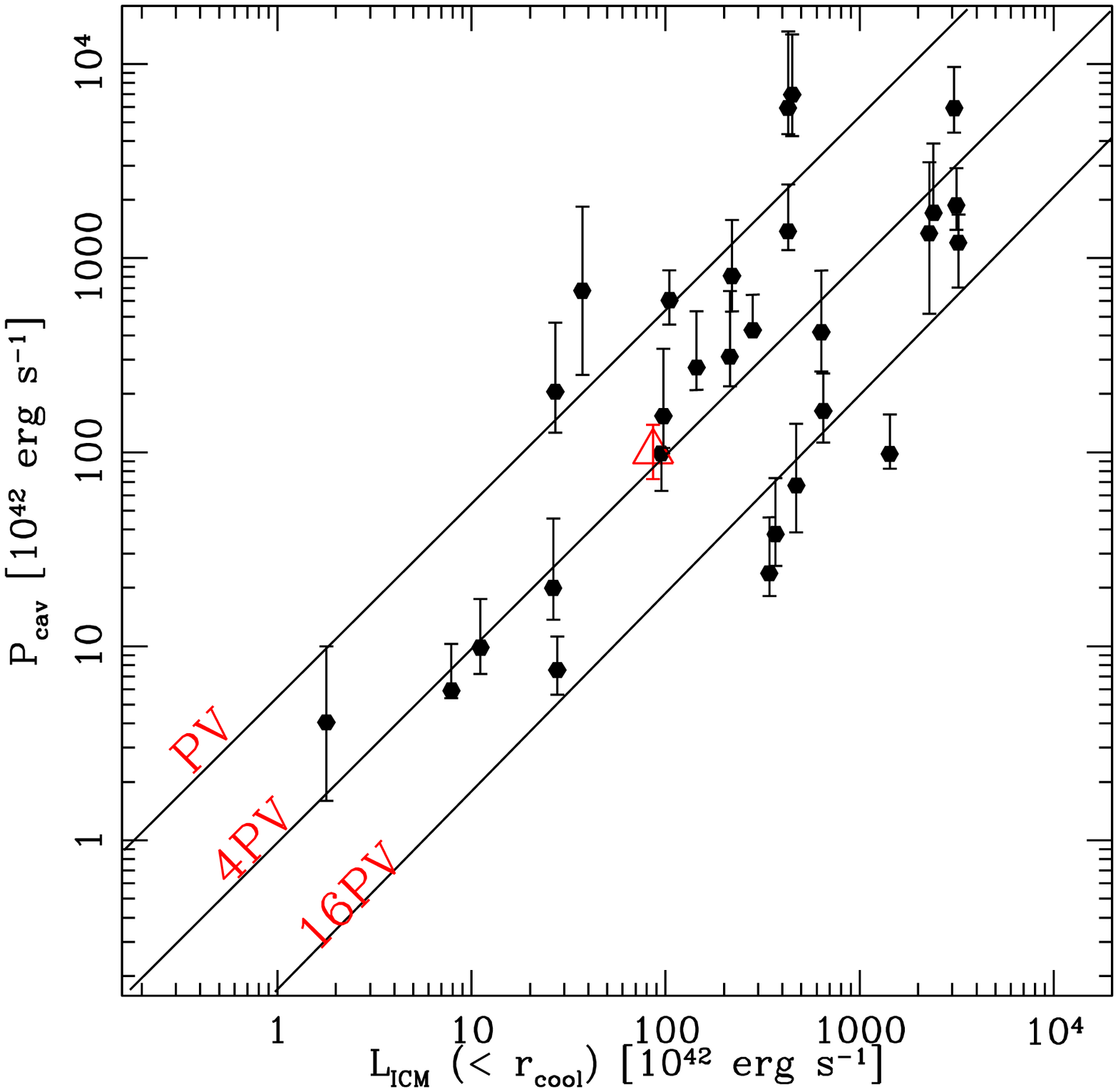}
\includegraphics[width=80mm,height=80mm]{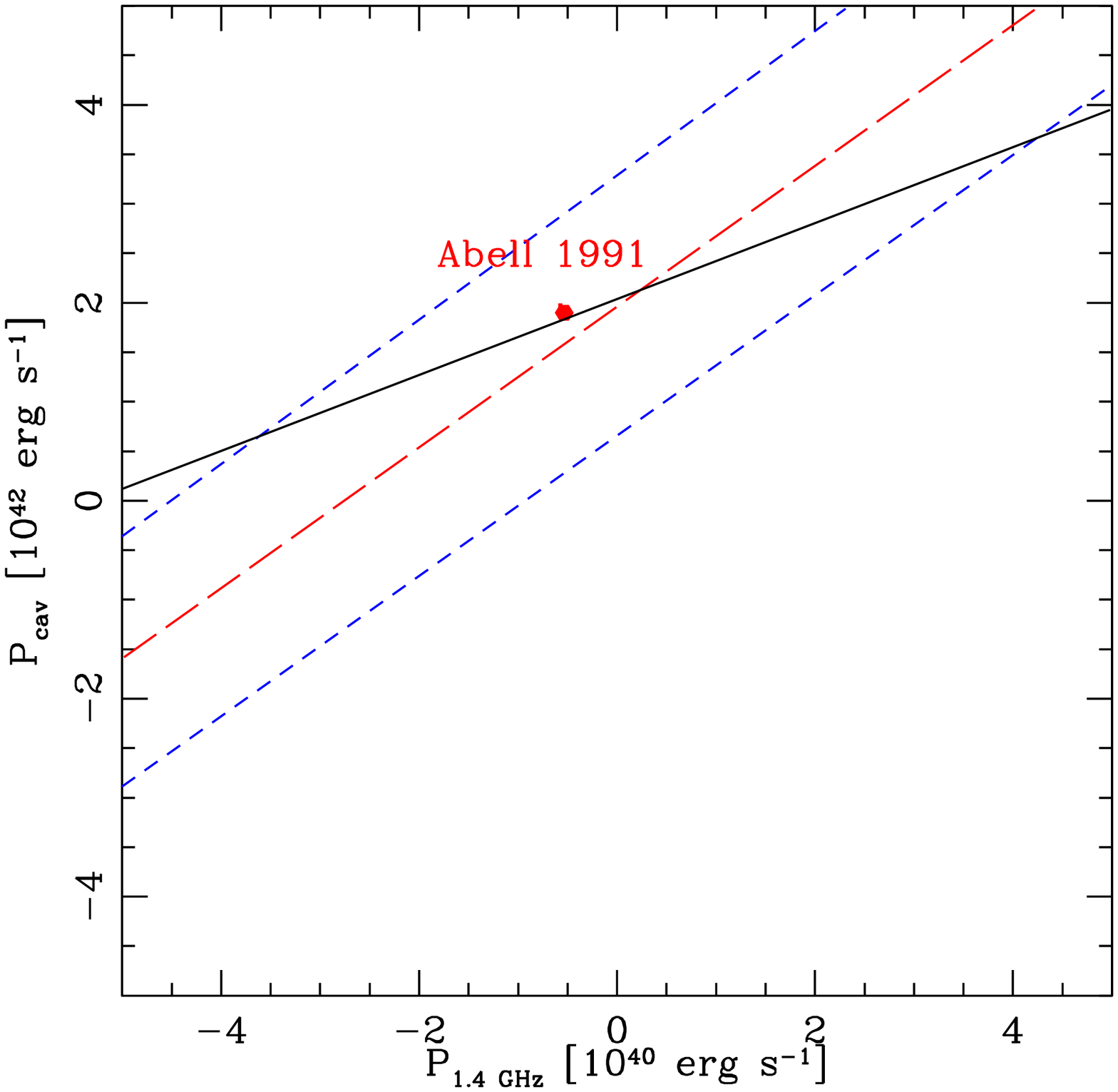}
}
\caption{\label{lx} {\it (left panel):} Cavity heating power vs the bolometric X-ray luminosity within cooling radius adopted from  \citet{2006ApJ...652..216R}. The diagonal lines denote cases where heating power equal cooling power ($P_{cav} = L_{ICM}$) for $pV$, 4$pV$ and 16$pV$ total enthalpy levels of the cavities from top to bottom. For comparison, we show a balance between the two for the case of Abell 1991 (open red triangle).  {\it (right panel):} Comparison of the balance between cavity heating power (P$_{cav}$) vs 1.4 GHz radio power for the sample studied by \citet{2010ApJ...720.1066C}. The red long dashed line represents the best fit relation for their sample of giant ellipticals (gEs), while the short dashed blue lines represent the upper and lower limits for their best fit relations. For comparison, we also show the best fit relation obtained  by \citet{2004ApJ...607..800B} for galaxy clusters (solid black line). Abell 1991 occupies position little above the best-fit line by \citet{2004ApJ...607..800B}. }
\end{figure*}

\subsection{Cavity Scaling Relations}
There are mounting evidences stating that the central dominant galaxies (CDGs) host radio sources \citep{2004ApJ...607..800B, 2006ApJ...652..216R,2010ApJ...720.1066C} and exhibit profound effect on the X-ray emitting ICM surrounding such systems. The ambipolar jets originating from the central source injects energy in to the ICM at regular intervals and hence results in the formation of under dense bubbles on the opposite sides of the nucleus which rises buoyantly in the cluster and expand until they reach pressure equilibrium with the surrounding ICM \citep{1974MNRAS.169..395B}. But whether the energy deposited by the central AGN is sufficient to account for the buoyant rise of the bubbles in the ICM  is not clear. 

With an objective to investigate association of this cooling flow galaxy with the central radio source, we have made use of the radio data available in the archive of the Very Large Array Faint Images of the Radio Sky at Twenty-Centimeters (VLA FIRST) survey. The radio emission contours derived from the analysis of radio data are overlaid on the unsharp masked X-ray image of Abell 1991 and is shown in Figure~\ref{ecimage} (d). This figure confirms the coincidence of a point-like source with the central dominant galaxy NGC 5778 and is consistent with that reported by \citet{2004ApJ...613..180S}, however, shows an offset of 10\arcsec relative to the X-ray peak of the cluster. Though this radio source shows an association with the S-cavity, no such association is evident for the case of N-cavity. Such cavities with no apparent spatial coincidence with the 1.4 GHz radio contours are referred as "ghost cavities" \citep{2000MNRAS.318L..65F,2001ApJ...562L.149M} and are thought to be formed by the previous outbursts. 

To investigate whether the energy deposited into the ICM from the central radio source is sufficient to account for the heating, we compare the mechanical power of the AGN with the cooling power of the ICM. AGN power, as discussed above, was estimated as the $pdV$ work required to inflate a cavity against the surrounding pressure of the hot ICM. For the case of Abell 1991, the total output power of the radio source derived from the measured values of pressure and volume of the X-ray cavities is found to be $\sim$ 8.64$\times$10$^{43}$\lum , while the X-ray luminosity within the cooling radius ($\sim$ 117 kpc) is found to be L$_{ICM} (<r_{cool}$)$\sim$ 6.04$\times$10$^{43}$\lum . Therefore, in this case, on the basis of simple energy arguments, it is evident that the radio source hosted by Abell 1991 on average is capable to deposit enough energy into the ICM so as to offset the cooling gas. We further check this balance by comparing its relevance with that of the sample studied by \cite{2006ApJ...652..216R}. For this we make use of their plot between the cavity power of the central AGN against total radiative luminosity of the intracluster gas within the cooling radius  (Figure~\ref{lx} \textit{left panel}). The thick diagonal lines in this figure represents the $P_{cav} = L_{ICM}$ for the total enthalpy levels equal to $pV$, 4$pV$ and 16$pV$ of the cavities. In this plot we indicate the Abell 1991 by red open triangle and is found to occupy a position on the 4$pV$ line. This in turn imply that the total power available within the cavities of Abell 1991 is sufficient to balance the entire radiative loss of the ICM within the cooling radius.  

\cite{2010ApJ...720.1066C} observed a strong power-law relationship between the mechanical power stored within the cavities and the 1.4 GHz radio power of the central source. We made use of this relationship to compare 1.4 GHz radio power ($P_{1.4~GHz}$) and cavity power ($P_{cav}$) of the Abell 1991 using the samples studied by \citet{2010ApJ...720.1066C} and \citet{2004ApJ...607..800B} and is shown in Figure~\ref{lx}~({\it right panel}). The red long dashed line represent the best fit relation of \cite{2010ApJ...720.1066C} for the sample of giant ellipticals (gEs), while small dashed blue lines denotes the upper and lower limits for their best fit relations. We also show the best-fit relation of \cite{2004ApJ...607..800B} for a sample of galaxy clusters studied by them and is shown by thick continuous line. In this plot Abell 1991 occupies a position confirming the correlation between the two and hence imply that the radio source hosted by this system has a capability to reverse the cooling flow. We have also compared the jet power (P$_{jet}$) of radio source  hosted by Abell 1991 with that of the observed radio power (P$_{radio}$) using the relation presented by \cite{2010ApJ...720.1066C} and found that, like other FR-I systems studied by \cite{2008MNRAS.386.1709C}, Abell 1991 also exhibit similar power capabilities.  Further, \cite{2012MNRAS.427.3468B} have demonstrated that every massive cluster hosts a radio source with its radio luminosity greater than 2.5$\times$10$^{30}$ erg s$^{-1}$ Hz$^{-1}$. Compared to this, the radio source associated with Abell 1991 have 1.4 GHz luminosity equal to $\sim$ 3.02$\times$10$^{30}$ erg s$^{-1}$ Hz$^{-1}$, and hence support the idea of self-regulating radio-mode outburst. Thus, all the attempts made here to relate the jet/cavity power with the X-ray/radio luminosity imply that the AGN outburts from the Abell 1991 are comparable to those needed for quenching the cooling flow. However, deep, high-resolution, multi-frequency radio observations of this system are called for arriving at proper conclusion. 
\section{ Summary}
\label{Summary.sec} 
In this paper we have presented results based on the re-analysis of the {\it Chandra} X-ray observations of the galaxy cluster Abell 1991 with an objective to investigate properties of X-ray cavities and the interplay with the central radio source. Summary of the main results is: 
\begin{enumerate}
\item Unsharp masked image as well as 2-d $\beta$ model subtracted residual image reveals a pair of ellipsoidal shaped X-ray cavities at a projected distance of about 12\,kpc from the central region of Abell 1991 and are consistent with those seen in several other cooling flow galaxies like NGC 6338. In addition to these prominent cavities, an excess X-ray emission region is also evident on the southern side of the S-cavity. Surface brightness profiles derived for this system revealed depressions in the X-ray emission at a distance of about 12\,kpc, confirming the presence of X-ray cavities.
\item Spectral analysis of X-ray photons extracted from both the cavities gives temperature of gas corresponding to the N and S cavities equal to $1.77\pm 0.15$ keV and 1.53$\pm$0.06 keV, respectively, while that of the excess X-ray emission region is 2.06$\pm$ 0.09 keV.
\item Like in other cooling flow clusters, temperature profile derived for Abell 1991 exhibits a positive temperature gradient, reaching to a maximum of 2.63 keV at $\sim$ 76 kpc and then declines in outward direction. 
\item  0.5$-$2.0 keV X-ray image of Abell 1991 reveals three different knot like structures in the central 15\arcsec region that are about 10\arcsec off the X-ray peak of the cluster. However, an additional absorption component was required to constrain combined spectrum of the knots.  
\item Total mechanical power of the cavity system in Abell 1991 is  P$_{cavity}$=8.64$\times$10$^{43}$ erg s$^{-1}$, while the luminosity of the gas within cooling radius (r$_{cool}$=117\,kpc) is L$_{cool} \sim$ 6.04$\times$10$^{43}$ erg s$^{-1}$. This is turn imply that the mechanical power injected by the central AGN through outbursts is sufficient to balance the radiative loss.
\item Radio emission map derived using VLA FIRST survey data shows an apparent coincidence with the S-cavity and  exhibit an offset of 10\arcsec relative to the X-ray peak of the Abell 1991.
\item AGN heating, as traced by the power content of the X-ray cavities, is capable of balancing the radiative losses of the ICM in this system. However, high resolution, low-frequency GMRT observations of Abell 1991 are required to examine the association of X-ray cavities with the radio-jets. 
\item We have detected a total of 21 point-like sources within the chip 7, majority of which are like 1.4 \Msun accreting neutron star binaries while few appears as heavily absorbed harder sources.
\end{enumerate}
\section*{Acknowledgments} 
The authors are grateful to the anonymous referee for their encouraging and constructive comments on the manuscript,  that greatly helped us to improve its quality. MBP gratefully acknowledges financial support from DST, New Delhi (file No. IF/10179), NDV acknowledges support from UGC, New Delhi (F.No. 36-240/2008-SR). Usage of the High Performance Computing facility procured under the DST-FIST scheme (SR/FST/PSI-145) is gratefully acknowledged. This work has made use of data from the \textit{Chandra} archive, NASA's Astrophysics Data System(ADS), Extragalactic Database (NED), and software provided by the Chandra X-ray Center (CXC).

\def\aj{AJ}%
\def\actaa{Acta Astron.}%
\def\araa{ARA\&A}%
\def\apj{ApJ}%
\def\apjl{ApJ}%
\def\apjs{ApJS}%
\def\ao{Appl.~Opt.}%
\def\apss{Ap\&SS}%
\def\aap{A\&A}%
\def\aapr{A\&A~Rev.}%
\def\aaps{A\&AS}%
\def\azh{AZh}%
\def\baas{BAAS}%
\def\bac{Bull. astr. Inst. Czechosl.}%
\def\caa{Chinese Astron. Astrophys.}%
\def\cjaa{Chinese J. Astron. Astrophys.}%
\def\icarus{Icarus}%
\def\jcap{J. Cosmology Astropart. Phys.}%
\def\jrasc{JRASC}%
\def\mnras{MNRAS}%
\def\memras{MmRAS}%
\def\na{New A}%
\def\nar{New A Rev.}%
\def\pasa{PASA}%
\def\pra{Phys.~Rev.~A}%
\def\prb{Phys.~Rev.~B}%
\def\prc{Phys.~Rev.~C}%
\def\prd{Phys.~Rev.~D}%
\def\pre{Phys.~Rev.~E}%
\def\prl{Phys.~Rev.~Lett.}%
\def\pasp{PASP}%
\def\pasj{PASJ}%
\def\qjras{QJRAS}%
\def\rmxaa{Rev. Mexicana Astron. Astrofis.}%
\def\skytel{S\&T}%
\def\solphys{Sol.~Phys.}%
\def\sovast{Soviet~Ast.}%
\def\ssr{Space~Sci.~Rev.}%
\def\zap{ZAp}%
\def\nat{Nature}%
\def\iaucirc{IAU~Circ.}%
\def\aplett{Astrophys.~Lett.}%
\def\apspr{Astrophys.~Space~Phys.~Res.}%
\def\bain{Bull.~Astron.~Inst.~Netherlands}%
\def\fcp{Fund.~Cosmic~Phys.}%
\def\gca{Geochim.~Cosmochim.~Acta}%
\def\grl{Geophys.~Res.~Lett.}%
\def\jcp{J.~Chem.~Phys.}%
\def\jgr{J.~Geophys.~Res.}%
\def\jqsrt{J.~Quant.~Spec.~Radiat.~Transf.}%
\def\memsai{Mem.~Soc.~Astron.~Italiana}%
\def\nphysa{Nucl.~Phys.~A}%
\def\physrep{Phys.~Rep.}%
\def\physscr{Phys.~Scr}%
\def\planss{Planet.~Space~Sci.}%
\def\procspie{Proc.~SPIE}%
\let\astap=\aap
\let\apjlett=\apjl
\let\apjsupp=\apjs
\let\applopt=\ao
\bibliographystyle{spr-mp-nameyear-cnd}
\bibliography{mybib}
\end{document}